\documentclass{iopart}

\eqnobysec

\begin{document}

\jl{1}

\title[Quantum crumpling]{The crumpling transition of membranes driven by quantum
fluctuations in a $D=\epsilon$ expansion}

\author{Georg Foltin}
\address{Institut f\"ur Theor\-etische
Physik, Hein\-rich\---Heine--Uni\-ver\-sit\"at D\"us\-sel\-dorf,
Uni\-ver\-sit\"ats\-strasse
1, D--40225 D\"ussel\-dorf, Germany}

\begin{abstract}
We consider a $D$-dimensional
fluid membrane in a $D+1$-dimensional embedding space, subject to
quantum fluctuations.
The corresponding action
is invariant under coordinate transformations and depends only on the shape
of the membrane and its variation, neglecting
tangential degrees of freedom.
We calculate the resulting field theory to one loop order in a
$D=\epsilon$ - expansion and find a quantum transition even at $T=0$.
\end{abstract}

\pacs{87.16.Dg, 05.30.-d, 68.35.Rh}


\section{Introduction}

Fluid membranes, like biomembranes, belong to the classical world---
there is no need to take quantum fluctuations into consideration
since typical temperatures are high and sizes are large.
Nevertheless, we might reduce temperature down to zero and increase $\hbar$
in a \textit{Gedankenexperiment} and study the quantum fluctuations of a
(non relativistic) flexible membrane. The most important feature of the resulting
model is a second-order quantum transition at a finite $\hbar$
from an almost flat to a crumpled phase, contrary to the thermal case, where
the membrane remains always in the crumpled phase \cite{Lei89}.
Ingredients of the action describing the quantum membrane are a kinetic energy
term, surface tension and the Canham-Helfrich curvature energy \cite{Can70}.
The latter elastic term disfavours curved configuration of the membrane and is
proportional to the square of the mean curvature, integrated over the surface of the membrane.
The action should not depend on the internal coordinates which are used to
parametrize the surface---it should be a functional of the geometry of the
membrane only.

Our model serves as a toy model for quantum
interfaces like the Helium liquid--vapour interface at very low temperatures
\cite{Urk25,Ese63}.
It is remarkable, that He$^3$ enriches at a He$^4$ interface and
acts as a surfactant,
lowering the surface tension of the He$^4$ interface \cite{Ese63}.
Therefore, the next-leading curvature terms become relevant.
The He$^3$ film on top of the He$^4$ bulk could be in fact a suitable
candidate for the quantum membrane under investigation.
It is, however, unlikey that the predicted second-order phase transition from a
flat to a crumpled interface (at finite $\hbar$) can be observed experimentally.
The surface tension of the He-system is still non-zero, and moreover, there is
no direct way to change $\hbar$ in an experiment. Nevertheless, it might be
possible by changing the He$^3$ concentration to tune the system closer to the
transition point.

So far, nonrelativistic quantum membranes with curvature stiffness
and surface tension were discussed in \cite{Bor00} at fixed
dimension $D=2$ of the membrane and $d=3$ of the embedding space and in
\cite{Bor99} for an infinite dimensional embedding space $d\rightarrow\infty$.
Extending these results we study the quantum fluctuations of a membrane
in a systematic $D=\epsilon$ expansion and find a fixed point
at zero temperature and finite $\hbar$ (for $D>0$), which is not seen by
\cite{Bor00}.

\section{The model}

Following \cite{Bor00} we start from the (imaginary time) action
\begin{equation}
\label{general}
S_0=\int \rmd t \rmd^D\sigma\sqrt{g}\left(\frac{1}{2\nu}(\partial_t\bi{X})^2+r+
\frac{2}{\alpha}H^2\right),
\end{equation}
where $\bi{X}(\sigma,t)$ describes the time-dependent
$D$-dimensional surface
embedded in a $D+1$-dimensional, Euclidean space. Hence $\sigma$ is a $D$-dimensional
set of internal coordinates which parametrize the surface.
$\rmd^D\sigma\sqrt{g}$ is the invariant element of area, $H$ is the mean 
curvature, $1/\nu$ is the mass density, $r$ is the surface tension and
$1/\alpha$ is the bending rigidity.
The action (\ref{general}) only makes sense in Lagrangian coordinates,
i.e. coordinates which follow the elements of fluid. Then,
$\bi{X}(\sigma,t)$ is (for fixed $\sigma$) the trajectory of an element
of fluid and $\partial_t\bi{X}$ the corresponding velocity. The velocity
might be decomposed into a normal part $v_\perp=
\bi{N}\cdot\partial_t\bi{X}$ and the tangential parts $u_i=\partial_i
\bi{X}\cdot\partial_t\bi{X}$, where $\bi{N}$ is the normal vector
and $\partial_i\bi{X}=\partial\bi{X}/\partial\sigma^i$ are the
tangential vectors (see e.~g. \cite{Kle89,Fol94}).
The normal velocity $v_\perp$ allows for a geometrical
interpretation---it encodes the variations of the shape of the membrane,
whereas the tangential velocity simply generates reparametrizations of the
surface (coordinate transformations).
Indeed, a purely tangential velocity field leaves the shape of the
membrane unchanged
and, therefore, belongs to degrees of freedom beyond the geometric
ones\footnote{
The divergence of the tangential velocity field, however,
is fixed by the (geometrical) condition $2Hv_\perp=D^iu_i$ in case of an
incompressible membrane.}.

Instead of using action (\ref{general}) and taking care of the tangential
degrees of freedom, we write down a simplified action, depending only
on the shape of the membrane and its variation (with conveniently chosen
coupling constants)
\begin{equation}
\label{our}
S_0/\hbar=\int\rmd t \rmd^D\sigma\sqrt{g}\left(\frac{\lambda}{2\hbar}v_\perp^2
+\frac{r}{\hbar}+\frac{2}{\hbar\lambda}H^2\right).
\end{equation}
It can be seen easily that $v_\perp$ is really a scalar under general
\textit{time-dependent} coordinate transformations
$\sigma\rightarrow\sigma(\sigma',t)$.
Consequently, the action $S_0$ itself is invariant under any time-dependent
coordinate transformation.
To calculate the partition function $\mathcal{Z}=\int D[\bi{X}]
\exp(-S_0/\hbar)$ and related expectation values, we have to sum over all
physically distinct surfaces $\bi{X}(\sigma,t)$ in a reparametrization-invariant
manner. Being far from trivial (see \cite{Cai94}, the quantum
case does not pose an extra complication), we have to restrict the
discussion of the measure problem to few remarks. At first, we
have to choose a certain representation of the surfaces (gauge fixing) in
order to avoid over-counting of surfaces with identical shapes, but different
coordinate systems. A common and practical choice is a representation of the
surface in terms of a (time dependent) heigth-variable $f(\mathbf{x},t)$ - the 
Monge representation ($\mathbf{x}$ is a $D-$dimensional Euclidean vector)
\begin{equation}
\label{monge}
\bi{X}(\mathbf{x},t)=\left(\mathbf{x},f(\mathbf{x},t)\right),
\end{equation}
which is connected to the Lagrangian
coordinates by a particular
time-dependent coordinate transformation. We convert the action (\ref{our}) into the Monge
representation using $v_\perp=\partial_t f/\sqrt{g}$ (where $g=1+(\nabla f)^2$)
and obtain
\begin{eqnarray}
\label{actionmonge}
S_0/\hbar&=&\int\rmd t\rmd^Dx\left[\frac{\lambda}{2\hbar}\frac{(\partial_t f)^2}
{\sqrt{g}}+\frac{r}{\hbar}\sqrt{g}\right.\nonumber\\
&&\left.+\frac{1}{2\hbar\lambda}\sqrt{g}\left(\nabla\cdot\left(\frac{1}
{\sqrt{g}}\nabla f\right)\right)^2\right]\\
&=&\int\rmd t\rmd^Dx\left[\frac{\lambda}{2\hbar}(\partial_t f)^2+\frac{r}{2\hbar}
\partial_i f\partial_i f+\frac{1}{2\hbar\lambda}(\partial^2 f)^2\right.
\nonumber\\
&&\mbox{}-\frac{\lambda}{4\hbar}(\partial_t f)^2\partial_i f\partial_i f
-\frac{r}{8\hbar}\partial_i f\partial_i f\partial_j f\partial_j f
\nonumber\\
&&\left.
\mbox{}-\frac{1}{4\hbar\lambda}\partial_i f\partial_i f(\partial^2 f)^2
-\frac{1}{\hbar\lambda}\partial_i f
\partial_j f\,\partial_i\partial_j f\,\partial^2 f\right]
+\Or ( f^6),\nonumber
\end{eqnarray}
where $i,j=1\ldots D$ \footnote
{The action (\ref{actionmonge}) differs from the one used in
\cite{Bor00} by a wrong sign in front of the first vertex.
The wrong sign, however, does not affect any consequence
drawn by \cite{Bor00} (for zero temperature),
since the authors study the field theory not at the lower critical dimension,
but at the dimension $D=2$,
where the flow of the coupling constants is mainly determined by their
naive dimensions.}.
The corresponding invariant measure $D[f]$ differs from the naive measure
$\propto\prod_{\mathbf{x}}\int\rmd f(\mathbf{x})$ by the so-called Fadeev-Popov
determinant and the Liouville term, which, however, contribute to two loop and
higher orders only. To one loop order, we may safely use the naive measure
instead \cite{Cai94}.
The lower critical dimension of the theory is
$D=0$, where the coupling constants in front of the kinetic energy and
in front of the curvature energy become marginal, as can be seen from the
dimensions of the coupling constants ($L$=length),
which are $\hbar\sim L^D$ and $r\sim L^{-2}$
($\lambda$ is rendered dimensionless, $t\sim L^2$).
Therefore, we have to calculate the quantum fluctuations of (\ref{actionmonge})
in a double $\hbar$ and $D=\epsilon$ expansion, which is done here to one
loop order with the help of dimensional regularization and the minimal
subtraction scheme in analogy with \cite{Wal79}.
The surface tension $r$ is a relevant parameter of the theory and is zero right at the critical
point. In fact, a non-zero $r$ imposes a finite correlation length $\xi=r^{-1/2}$ on the
propagator of action (\ref{actionmonge}).
Not included in the action (\ref{actionmonge}) are the integral over the
scalar curvature $R$
(which does not yield a contribution to one loop order and which is a
topological invariant for $D=2$) and a boundary term---the cross term
$2\int\rmd t\rmd^Dx\sqrt{g}\,v_\perp H=-\int\rmd t \partial_t A$,
where $A$ is the surface area.

\section{Field theory}

The bare $T=0$ two-point vertex function $\Gamma_{0,2}(q,\omega)$ reads
(denoting from now on bare quantities with a subscript 0)
\begin{eqnarray}
\Gamma_{0,2}(q,\omega)&=&\frac{\lambda_0}{\hbar_0}\omega^2+
\frac{r_0}{\hbar_0}q^2+\frac{1}{\hbar_0\lambda_0}q^4\nonumber\\
&&\mbox{}+\frac{\omega^2}{2}\lambda_0(r_0\lambda_0)^{\epsilon/2}I_\epsilon
-\frac{q^2}{2+\epsilon}r_0(r_0\lambda_0)^{\epsilon/2}I_\epsilon
\nonumber\\
&&\mbox{}+q^4\left(\frac{1}{2}+\frac{2}{\epsilon}\right)\frac{1}{\lambda_0}
(r_0\lambda_0)^{\epsilon/2}I_\epsilon,
\end{eqnarray}
where $I_\epsilon=(4\pi)^{-\epsilon/2-1/2}\Gamma(1-\epsilon/2)\Gamma(\epsilon/2
+1/2)/\Gamma(1+\epsilon/2)$ ($I_\epsilon$ is finite in the limit $\epsilon
\rightarrow 0$).
We introduce renormalized couplings $\hbar,\lambda, r$
by $\hbar_0=\mu^{-\epsilon}\hbar Z_\hbar/I_\epsilon$, $r_0=Z_r r$ and
$\lambda_0=Z_\lambda\lambda$ ($\mu$ is an arbitrary momentum scale)
and require the renormalized vertex function
\begin{eqnarray}
\Gamma_2(q,\omega)&=&\frac{\mu^\epsilon I_\epsilon}{\hbar}\left(
\frac{Z_\lambda}{Z_\hbar}\lambda\omega^2+\frac{Z_r}{Z_\hbar}rq^2+\frac{1}
{\lambda Z_\lambda Z_\hbar}q^4\right.\nonumber\\
&&\left.\mbox{}+\frac{\omega^2}{2}\lambda\hbar Z_\lambda-\frac{q^2}{2+\epsilon}
r\hbar Z_r+q^4\left(\frac{1}{2}+\frac{2}{\epsilon}\right)\frac{\hbar}
{\lambda Z_\lambda}\right)
\end{eqnarray}
to be finite in the limit $\epsilon\rightarrow 0$. We find $Z_\hbar=Z_\lambda
=Z_r=1+\hbar/\epsilon+\Or (\hbar^2)$.
The RG-equation is readily derived from the fact, that the bare quantities
do not depent on the momentum scale $\mu$, yielding
\begin{equation}
\left(
\mu\partial_\mu+\beta(\hbar)\partial_\hbar+\gamma r\partial_r +\zeta\lambda
\partial\lambda\right)\Gamma_2(q,\omega)=0
\end{equation}
with the beta-function $\beta(\hbar)=\hbar(\epsilon-\hbar)$.
The theory has an ultraviolet stable fixed point (not seen by
\cite{Bor00}) $\hbar^*=\epsilon$ which
corresponds to a quantum-phase transition at a \textit{finite}
$\hbar$ (for $D>0$ and $T=0$) from a smooth phase for small $\hbar$ to
a crumpled phase for $\hbar>\hbar^*$.
The main effect of quantum fluctuations on the statistics of membranes
is a shift of the lower critical dimension from $D=2$ to $D=0$.
For slightly larger  dimensions a quantum-phase transition takes place at
a finite $\hbar^*$. We expect that this picture remains true up to the
physical dimension $D=2$, although the numerical accuracy for the one-loop
critical exponents (not presented in this short communication)
should be quite limited.

\ack

I thank R Bausch, H K Janssen and R Blossey
for useful discussions. This work has been supported by the Deutsche
Forschungsgemeinschaft under SFB~237.

\appendix

\section{Details of the calculation}

The bare two-point vertex function reads
\begin{eqnarray}
\Gamma_{0,2}(q,\omega)&=&\frac{\lambda}{\hbar}\omega^2+
\frac{r}{\hbar}q^2+\frac{1}{\hbar\lambda}q^4\nonumber\\
&&\mbox{}-\frac{\lambda}{2\hbar}\omega^2\,I_2-\frac{\lambda}{2\hbar}q^2\,I_1
-\frac{r}{2\hbar}q^2\,I_2-\frac{r}{\hbar\epsilon}q^2\,I_2
\nonumber\\
&&\mbox{}-\frac{1}{2\hbar\lambda}q^2\,I_3-\frac{1}{2\hbar\lambda}q^4\,I_2
-\frac{2}{\hbar\lambda\epsilon}q^2\,I_3
-\frac{2}{\hbar\lambda\epsilon}q^4\,I_2
\end{eqnarray}
with the Feynman-integrals
\begin{eqnarray}
I_1&=&\hbar\int\frac{\rmd\omega}{2\pi}\,\frac{\rmd^\epsilon q}{(2\pi)^\epsilon}\,
\frac{\omega^2}{\lambda\omega^2+rq^2+\lambda^{-1}q^4},\nonumber\\
I_2&=&\hbar\int\frac{\rmd\omega}{2\pi}\,\frac{\rmd^\epsilon q}{(2\pi)^\epsilon}\,
\frac{q^2}{\lambda\omega^2+rq^2+\lambda^{-1}q^4},\nonumber\\
I_3&=&\hbar\int\frac{\rmd\omega}{2\pi}\,\frac{\rmd^\epsilon q}{(2\pi)^\epsilon}\,
\frac{q^4}{\lambda\omega^2+rq^2+\lambda^{-1}q^4}.
\end{eqnarray}
Within dimensional regularization we have ($\int\rmd\omega\rmd^\epsilon q\equiv0$)
\begin{eqnarray}
\label{i1}
I_1&=&\frac{\hbar}{\lambda}
\int\frac{\rmd\omega}{2\pi}\,\frac{\rmd^\epsilon q}{(2\pi)^\epsilon}\,
\frac{\lambda\omega^2+rq^2+\lambda^{-1}q^4-rq^2-\lambda^{-1}q^4}
{\lambda\omega^2+rq^2+\lambda^{-1}q^4}\nonumber\\
&=&\mbox{}-\frac{r}{\lambda}\,I_2-\frac{1}{\lambda^2}\,I_3.
\end{eqnarray}
To evaluate $I_2$, we substitute $\omega\rightarrow\omega q^2$ and find
\begin{equation}
I_2=\hbar\int\frac{\rmd^\epsilon q}{(2\pi)^\epsilon}\,\frac{\rmd\omega}{2\pi}\,
\frac{q^2}{\lambda\omega^2q^2+r+\lambda^{-1}q^2}.
\end{equation}
Now we are able to perform the $q$-integration
\begin{equation}
I_2=-\frac{\hbar}{r}\int\frac{\rmd\omega}{2\pi}\left(\frac{r}{\lambda\omega^2+
\lambda^{-1}}\right)^{1+\epsilon/2}\,I,
\end{equation}
where
\begin{equation}
I=\int\frac{\rmd^\epsilon k}{(2\pi)^\epsilon}\,\frac{-k^2}{k^2+1}=\frac{1}
{(4\pi)^{\epsilon/2}}\Gamma(1-\epsilon/2).
\end{equation}
With the help of $(2\pi)^{-1}\int\rmd s(s^2+1)^{-1-\epsilon/2}=(4\pi)^{-1/2}
\Gamma(\epsilon/2+1/2)/\Gamma(1+\epsilon/2)$ we obtain
\begin{equation}
I_2=-\hbar(r\lambda)^{\epsilon/2}I_\epsilon,
\end{equation}
where
\begin{equation}
I_\epsilon=\frac{1}{(4\pi)^{\epsilon/2+1/2}}
\frac{\Gamma(1-\epsilon/2)\Gamma(\epsilon/2
+1/2)}{\Gamma(1+\epsilon/2)}.
\end{equation}
An analogous calculation yields
\begin{equation}
\label{i3}
I_3=-\lambda r\frac{1+\epsilon}{2+\epsilon}I_2=\hbar\lambda r
(r\lambda)^{\epsilon/2}\frac{1+\epsilon}{2+\epsilon}I_\epsilon.
\end{equation}

Within a cut-off regularization scheme instead of dimensional regularization
additional divergent terms show up in equation (\ref{i1}) and equation (\ref{i3})
which can be absorbed by an additive renormalization of the surface tension $r$.

\end{document}